\documentclass[twocolumn]{aastex62}

\usepackage{natbib}
\usepackage{amsmath}
\usepackage{color}
\usepackage{multirow}

\usepackage{rotating}

\usepackage{hyperref}
\hypersetup{
    colorlinks=true,
    linkcolor=blue,
    citecolor=blue,
    filecolor=magenta,      
    urlcolor=cyan,
}

\defcitealias{Ho2017}{Paper~I}
\defcitealias{Martin2019}{M19}

\shorttitle{3D Disk Orientation and Gas Inflows}
\shortauthors{Ho \& Martin}

\begin{document}
%
%
\def\etali{{\it et al.\thinspace}}
\def\etns{{\rm et\thinspace al.}}   

\def\Heff{$H_{\mathrm{eff}}$}
\def\hv{$h_v$}

\def\mgIIdb{\ion{Mg}{2} $\lambda\lambda$2796, 2803}
\def\mgIIdbl{\ion{Mg}{2} $\lambda$2796}
\def\mgIIdbu{\ion{Mg}{2} $\lambda$2803}
\def\oI{[\ion{O}{1}] $\lambda$6300}
\def\nII{[\ion{N}{2}] $\lambda\lambda$6548, 6583}
\def\sII{[\ion{S}{2}] $\lambda\lambda$6716, 6731}
\def\oIII{[\ion{O}{3}] $\lambda$5007}
\def\hI{\mbox {\ion{H}{1}}}

\def\kms{\mbox{km s$^{-1}$}}
\def\kmstb{km s$^{-1}$}
\def\micron{\mbox{$\mu$m}}

%
%
\def \dlow {\mbox{$400 {\rm ~l~mm}^{-1}$}}
\def \dhigh {\mbox{$600 {\rm ~l~mm}^{-1}$}}
\newcommand{\be}{\begin{equation}} \newcommand{\ba}{\begin{eqnarray}}
\newcommand{\ee}{\end{equation}} \newcommand{\ea}{\end{eqnarray}}
\def\etal{{\it et al.\thinspace}}
\def\-{{\em{---}}}
\def \mA {\mbox{${\rm m \AA} $} }
\def \rr {\mbox{${\rm RR}$} }
\def \rarb {\mbox{${\rm R_AR_B}$} }
\def \rara {\mbox{${\rm R_AR_A}$} }
\def \dd {\mbox{${\rm DD}$} }
\def \dada {\mbox{${\rm D_AD_A}$} }
\def \dadb {\mbox{${\rm D_AD_B}$} }
\def \dr {\mbox{${\rm DR}$} }
\def \darb {\mbox{${\rm D_AR_B}$} }
\def \dara {\mbox{${\rm D_AR_A}$} }
\def \dbra {\mbox{${\rm D_BR_A}$} }
\def \hMpc      {h^{-1}{\rm\ Mpc}}
\def \hkpc      {h^{-1}{\rm\ kpc}}
\def \h         {\hbox{$\, h$} }
\def \hinv      {\hbox{$\, h^{-1}$} }
\def \hinvseven    {\hbox{$\, h_{70}^{-1}$} }
\def\ewr{\mbox {EW$_r$}}
\def\ewo{\mbox {EW$_o$}}
\def\H7{\mbox {$h_{0.7}$}}
\def\naI{\mbox {\ion{Na}{1}}}
\def\mgI{\mbox {\ion{Mg}{1}}}
\def\feI{\mbox {\ion{Fe}{1}}}
\def\oVI{\mbox {\ion{O}{6}}}
\def\znII{\mbox {\sc Zn~II~}}
\def\crII{\mbox {\sc Cr~II~}}
\def\alI{\mbox {\sc Al~I~}}
\def\alII{\mbox {\sc Al~II~}}
\def\alIII{\mbox {\sc Al~III~}}
\def\mgII{\mbox {\ion{Mg}{2}}}
\def\mnII{\mbox {\ion{Mn}{2}}}
\def\niII{\mbox {\ion{Ni}{2}}}
\def\feII{\mbox {\ion{Fe}{2}}}
\def\feIII{\mbox {\ion{Fe}{3}}}
\def\cIV{\mbox {\ion{C}{4}}}
\def\sV{\mbox {\ion{S}{5}}}
\def\siIV{\mbox {\ion{Si}{4}}}
\def\siII{\mbox {\ion{Si}{2}}}
\def\siI{\mbox {\ion{Si}{1}}}
\def\cII{\mbox {\ion{C}{2}}}
\def\cIII{\mbox {\ion{C}{3}}}
\def\llambda{\mbox {$\lambda$}}
\def\mstar{\mbox {$M_*$}}
\def\hlen{\mbox {$h_{0.7}^{-1}$}}
\def\lstarlya{\mbox {$L^*_{Ly\alpha}$}}
\def\IZw18{I~Zw~18}
\def\m82{M82}
\def\Ab{Abell~}
\def\gi{\mbox {\rm g-i}}
\def\ug{\mbox {\rm u-g}}
\def\br{\mbox {\rm b-r}}
\def\eqn{equation}
\def\vesc{\mbox {$v_{\rm esc}$}}
\def\heha{\mbox {He~I~$\lambda 5876$ / H$\alpha$}}
\def\xhe{\mbox {$\chi({\rm He}) / \chi({\rm H})$} }
\def\heii{\mbox {${\rm He}^+$}}
\def\he{\mbox {\rm He}}
\def\hii{\mbox {${\rm H}^+$}}
\def\h{\mbox {\rm H}}
\def\mab{\mbox {$\rm m_{AB}$}}
\def\ssp{\baselineskip=13pt plus 1pt minus 1pt}
\def\tsp{\baselineskip=5pt plus 1pt minus 1pt}
%
%
\def\deg{\mbox {$^{\circ}$}}
\def\msun{\mbox {${\rm ~M_\odot}$}}
\def\zsun{\mbox {${\rm ~Z_{\odot}}$}}
\def\lsun{\mbox {${~\rm L_\odot}$}}
\def\msunyr{\mbox {$~{\rm M_\odot}$~yr$^{-1}$}}
\def\angs{\mbox {~\AA}}
\def\lya{\mbox {Ly$\alpha$}}
\def\lyb{\mbox {Ly$\beta$}}
\def\Ha{\mbox {H$\alpha$}}
\def\Hb{\mbox {H$\beta$}}
\def\Hg{\mbox {H$\gamma$}}
\def\tion{\mbox {$T_{\rm ion}$~}}
\def\ch{\mbox {$\bigtriangleup$}}
\def\grad{\mbox {$\bigtriangledown$}}
\def\lstar{\mbox {$L^*$}}
\def\line{\mbox {~$\lambda$}}
\def\lines{\mbox {~$\lambda\lambda$~}}
\def\h0{\mbox {~H$_0$}}
\def\q0{\mbox {~q$_0$}}
%
%
\def\auroral{[OIII]~$\lambda4363$~}
\def\auroral{[OIII]~$\lambda4363$~}
\def\ohsun{\mbox {(O/H)$_{\odot}$~}}

\def\O1ha{[OI]$\lambda6300$~/~H$\alpha$~}
\def\Ru{[OII]$\lambda\lambda3727$~/~[OIII]$\lambda5007$~}
\def\s2ha{[SII]$\lambda\lambda6717,31$~/~H$\alpha$~}
\def\2z2{HeII~$\lambda4686$~}
\def\z7{[NII]~$\lambda6583$ }
\def\N2{[NII]~$\lambda6583$~/~H$\alpha$~}
\def\16z2{[SII]~$\lambda\lambda6717, 6731$ }
\def\HgI{HgI~$\lambda4358$~}
\def\Sdensity{[SII]~$\lambda6717 / \lambda6731$}
\def\Temp{[OIII]~$\lambda\lambda4959 + 5007 ~{\rm to}~ \lambda4363$~}
%
%
\def\j{J}
\def\n{NGC~}
\def\asec{\ifmmode {'' }\else $''~$\fi}  
\def\amin{\ifmmode {' }\else $'~$\fi}    
\def\arcsper{\ifmmode \rlap.{'' }\else $\rlap{.}'' $\fi} 
\def\arcmper{\ifmmode \rlap.{' }\else $\rlap{.}' $\fi} 
\def\sles{\lesssim}
\def\sgreat{\gtrsim}
%
%
\def\gapp{\mbox {$_>\atop{^\sim}$}}  
\def\lapp{\mbox {$_<\atop{^\sim}$}}  
%
\def\kms{\mbox {~km~s$^{-1}$}}
\def\ergsec{~ergs~s$^{-1}$~}
\def\sb{~ergs~s$^{-1}$~cm$^{-2}$~arcsec$^{-2}$}
\def\flux{~ergs~s$^{-1}$~cm$^{-2}$}
\def\flam{~ergs~s$^{-1}$~cm$^{-2}$ \AA$^{-1}$}
\def\cm3{~cm$^{-3}$}
\def\col{\mbox {~cm$^{-2}$}}
\def\mpc3{~Mpc$^{3}$}
\def\mpc-3{~Mpc$^{-3}$}
\def\rate{~sec$~{-1}$}
\def\um{~${\mu}$m~}
\def\fig{{Figure}}
\def\figs{{Figures}}
\def\tbl{{Table}~}
\def\sec{{Sec.}~}
\def\x{{X-ray}~}
\def\xs{{X-rays}~}
\def\X{{X-Ray}~}

%
\def\et{{\rm et\thinspace al.}\ }   
\def\ets{{\rm et\thinspace al.'s}\ }   
\def\reff{\par\noindent\parskip=1pt\hangindent=3pc\hangafter=1}
%
%

%
\def\beginrefs{
         {\normalsize}
         {\noindent}
         \small
        \baselineskip=11pt
        \parindent=0pt
        \frenchspacing
        \parskip=1pt plus 1pt
        \everypar={\hangindent=0.42in}}

\title{Resolving 3D Disk Orientation using High-Resolution Images: 
New Constraints on Circumgalactic Gas Inflows
}

\email{shho@physics.tamu.edu, cmartin@physics.ucsb.edu}

\author{Stephanie H. Ho}
\affiliation{George P.~and Cynthia Woods Mitchell Institute for Fundamental Physics and Astronomy, Texas A\&M University, College Station, TX 77843-4242, USA}
\affiliation{Department of Physics and Astronomy, Texas A\&M University, College Station, TX 77843-4242, USA}
\affiliation{Department of Physics, University of California, Santa Barbara, CA 93106, USA}

\author{Crystal L. Martin}
\affiliation{Department of Physics, University of California, Santa Barbara, CA 93106, USA}



\begin{abstract}


We constrain gas inflow speeds in 
star-forming galaxies
with color gradients consistent with inside-out disk growth.  
Our method combines new measurements of disk orientation 
with previously described circumgalactic absorption 
in background quasar spectra.  
Two quantities, 
a position angle and an axis ratio, 
describe the projected shape of each galactic disk on the sky, 
leaving an ambiguity about which side of the minor axis is
tipped toward the observer.  
This degeneracy regarding the 3D orientation of disks
has compromised previous efforts to measure gas inflow speeds.  
We present HST and Keck/LGSAO
imaging that resolves the spiral structure in 
five galaxies at redshift $z\approx0.2$.
We determine the sign of the disk inclination
for four galaxies, 
under the assumption that spiral arms trail the rotation.  
We project models for both radial infall in the disk plane 
and circular orbits onto each quasar sightline.  
We compare the resulting line-of-sight velocities 
to the observed velocity range of \mgII\ absorption in 
spectra of background quasars, 
which intersect the disk plane at radii between 69 and 115 kpc.
For two sightlines,
we constrain the maximum radial inflow speeds as 30-40\kms.   
We also rule out a velocity component 
from radial inflow in one sightline, 
suggesting that the structures feeding gas to 
these growing disks
do not have unity covering factor.
We recommend appropriate selection criteria 
for building larger samples of galaxy--quasar pairs 
that produce orientations
sensitive to constraining inflow properties.

\end{abstract}


\section{Introduction}
\label{sec:intro}

Gas accretion onto galaxies shapes the growth of their disks.
Decades of observations have  demonstrated 
that galaxies need a continuous gas supply 
to explain the star formation history 
and the stellar metallicity distribution of the disks.
Without a continuous gas supply,
the gas reservior around galaxies will exhaust within a few gigayears,
and the galaxies cannot sustain their star formation rates
\citep{Bigiel2008,Bigiel2011, Leroy2008,Leroy2013,Rahman2012}.  
The accreting gas will thereby prolong
the gas consumption time \citep{Kennicutt1983}
and explain the color of galaxy disks 
along the Hubble sequence \citep{Kennicutt1998}.  
The infall of metal-poor gas also explains
the relative paucity of low metallicity stars in the disk,
known as the G-dwarf problem in the solar neighborhood
\citep{vandenBergh1962,Schmidt1963,SommerLarsen1991}
but not unique to the Milky Way \citep{Worthey1996}.

In hydrodynamical simulations,
galaxies accrete cooling gas to grow the galactic disks
\citep{Oppenheimer2010,Brook2012,Shen2012,Ford2014,Christensen2016}.
The torques generated by the disk align the cooling, infalling gas with 
the pre-existing disk \citep{Danovich2012,Danovich2015}. 
The newly accreted gas then forms an extended cold flow disk 
that corotates with the central disk \citep{Stewart2011b,Stewart2013,Stewart2017}.  
As gas accreted at later times has higher angular momentum,
the late time infall builds the disk inside-out
\citep{Kimm2011,Pichon2011, Lagos2017,ElBadry2018}.

Direct observations of gas accretion onto galaxies
remain sparse \citep{Putman2012}.
Nevertheless, 
recent studies of the circumgalactic medium (CGM) 
through quasar sightlines
have shed light on detecting the inflowing gas.
The CGM extends to the galaxy virial radii \citep{Tumlinson2017}
and contains 
a significant fraction of the baryonic mass
associated with the galaxy halos \citep{Werk2014}.
For the low-ionization-state gas (e.g., Mg$^+$) 
detected as intervening absorption
in quasar sightlines,
absorption near the galaxy major and minor axes is 
often explained by inflows and outflows
(e.g., \citealt{Bouche2012,Kacprzak2012ApJ,Kacprzak2015,Nielsen2015});
minor- and major-axis sightlines
intersect winds blown out perpendicular to the disk plane
and gas accreted near the disk plane, respectively \citep{Shen2012}.
In addition, 
\citet[hereafter \citetalias{Martin2019}]{Martin2019}
studied the circumgalactic gas kinematics
of a sample of 50 $z\approx0.2$ blue galaxies,
and the galaxies have 
quasar sightlines intersecting the inner CGM
at all azimuthal angles.\footnote{
    We define the azimuthal angle
    as the angle between
    the galaxy major axis and
    the line joining the quasar and the galaxy center.}
They found that the \mgII\ absorption strength 
increases with the velocity range,
both of which increase toward the minor axis.
They concluded that 
minor-axis sightlines intersect
the CGM kinematically disturbed by galactic outflows,
a property that has been shown for galaxies
undergoing or have recently undergone a strong starburst 
\citep{Heckman2017}.

\citet[hereafter \citetalias{Ho2017}]{Ho2017}
studied a subset of 15 galaxy--quasar pairs 
from \citetalias{Martin2019},
and all 15 quasar sightlines lie within 30\deg\ from 
the galaxy major axes.  
In 13 out of the 15 major-axis sightlines, 
Ho et al.\ detected \mgII\ absorption, 
and the \mgII\ Doppler shift shares the same sign as 
the galactic disk rotation at the quasar side of the galaxy.  
This implies that the \mgII\ gas in the inner CGM
corotates with the galaxy disk.  
Moreover, \citetalias{Martin2019} found no 
net counter-rotating \mgII\ systems
within 45\deg\ of the major axis.
Therefore,
these results have strengthened previous results
from smaller, diverse galaxy samples that
also found corotating low-ionization-state gas in 
major-axis sightlines 
\citep{Steidel2002,Kacprzak2010,Bouche2013,Bouche2016,Zabl2019},
a property that also describes Ly$\alpha$ absorbers
\citep{Barcons1995,Prochaska1997,Chen2005}.

To explain the kinematics of the \mgII\ absorbing gas, 
\citetalias{Ho2017} explored 
simple models to describe the velocity range 
spanned by the absorption.  
We have adopted disk models,
instead of rotating halo models, 
as the simplest configuration.  
This is motivated in part by hydrodynamic simulations
that show the corotation between
the central disks and the extended cold gas disks
(e.g., \citealt{Stewart2011b,ElBadry2018}).
Previous observational work
has also attempted to 
model the Doppler shifts and/or velocity ranges
of Ly$\alpha$ and \mgII\ absorption systems
with disk rotation
(e.g., \mgII\ in \citealt{Steidel2002}
and \citealt{Kacprzak2010,kacprzak2011ApJ},
and Ly$\alpha$ in 
\citealt{Barcons1995,Prochaska1997,Chen2005}).  
It is well known that thin, rotating disks fail to
reproduce the broad velocity range of absorption systems.  
A wider velocity range is produced by 
thick disks (cylinders), 
a lag in rotation (i.e., a vertical velocity gradient),
radial inflow on the disk plane, 
or a combination of these options
\citep{Steidel2002,Kacprzak2010,kacprzak2011ApJ,Ho2017}. 
We have selected major-axis sightlines
that intersect the inner CGM of galaxies with inclined disks.
If the gas disks extend beyond the visible disks of galaxies,
then our sightlines will intersect gas  
near the extended disk planes 
within half the virial radii.
Therefore, motivated by previous work
and our selection of major-axis sightlines,
we adopt a disk geometry 
and explore the consequences for the gas dynamics.

\begin{figure}[htb]
    \centering
    \includegraphics[width=1.0\linewidth]{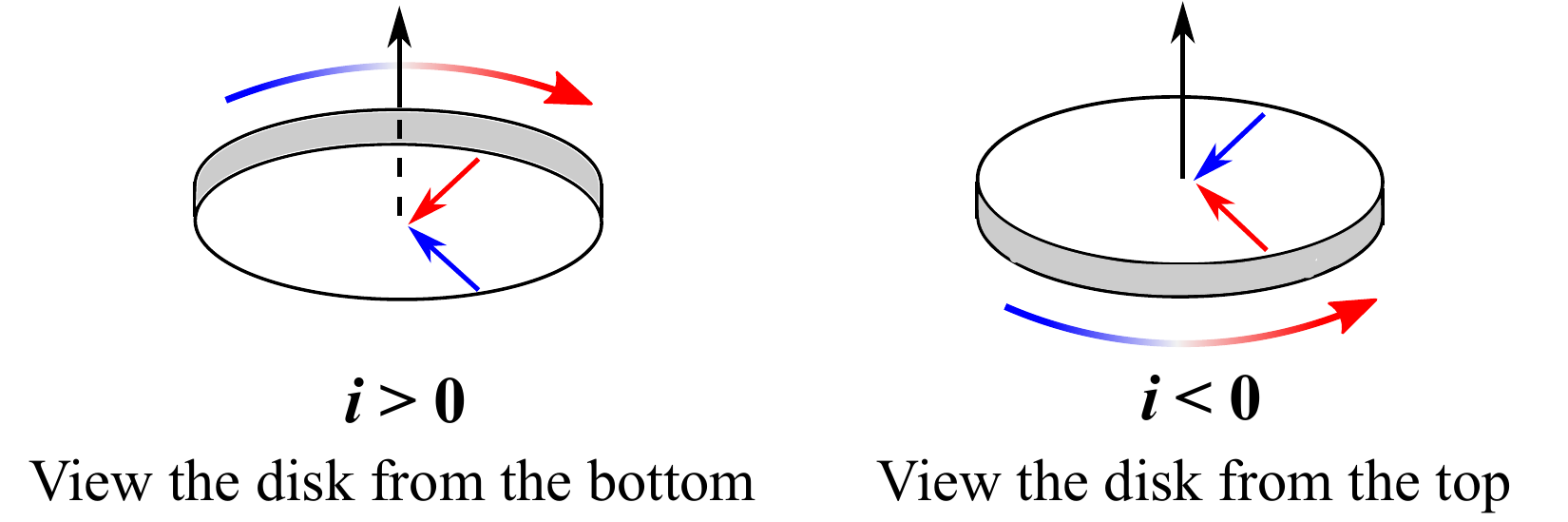}
    \caption{
        Sign of disk inclination.  
        For a measured disk axis-ratio, 
        the disk can tilt in two different ways
        with respect the plane of the sky.  
        From the perspective of the observer, 
        switching the sign of disk inclination
        flips the near side and the far side of the galaxy disk.
        }
    \label{fig:disk_flip} 
\end{figure}

With a disk geometry, 
the modeled line-of-sight (LOS) velocity 
depends on the galaxy disk orientation in 3D-space.
In addition to the disk axis-ratio and position angle,
establishing the 3D disk orientation 
requires the sign of disk inclination. 
The sign of disk inclination indicates
how the galaxy disk tilts with respect to the plane of the sky.    
From the perspective of the observer, 
switching the sign 
flips the near side and the far 
side of the galaxy disk;  
Figure~\ref{fig:disk_flip} illustrates this concept.
We have implemented this geometrical difference into 
our inflow model in \citetalias{Ho2017}.  
The model combines a tangential (rotation) 
and a radial (inflow) velocity component,
producing gas spiraling inward on the disk plane. 
When the disk flips,
the tangential and radial velocity components add together differently.  
This alters whether the radial inflow boosts or cancels 
the projected rotation velocity, 
resulting in different modeled LOS velocity ranges.
We have explained the model in detail and demonstrated this asymmetry 
in the Appendix of \citetalias{Ho2017}.

%
In general, 
the asymmetry originates from
velocity vectors added together differently when the disk flips.
Through the simple radial inflow model,
we will demonstrate the use of the disk tilt 
to constrain disk inflows.  
In \citetalias{Martin2019},
we have also discussed how the disk tilt 
constrains outflow models 
and explains minor-axis absorption.  
Hence, the disk tilt 
is an important parameter when
modeling circumgalactic gas kinematics.

\begin{figure}[htb]
    \centering
    \includegraphics[width=0.9\linewidth]{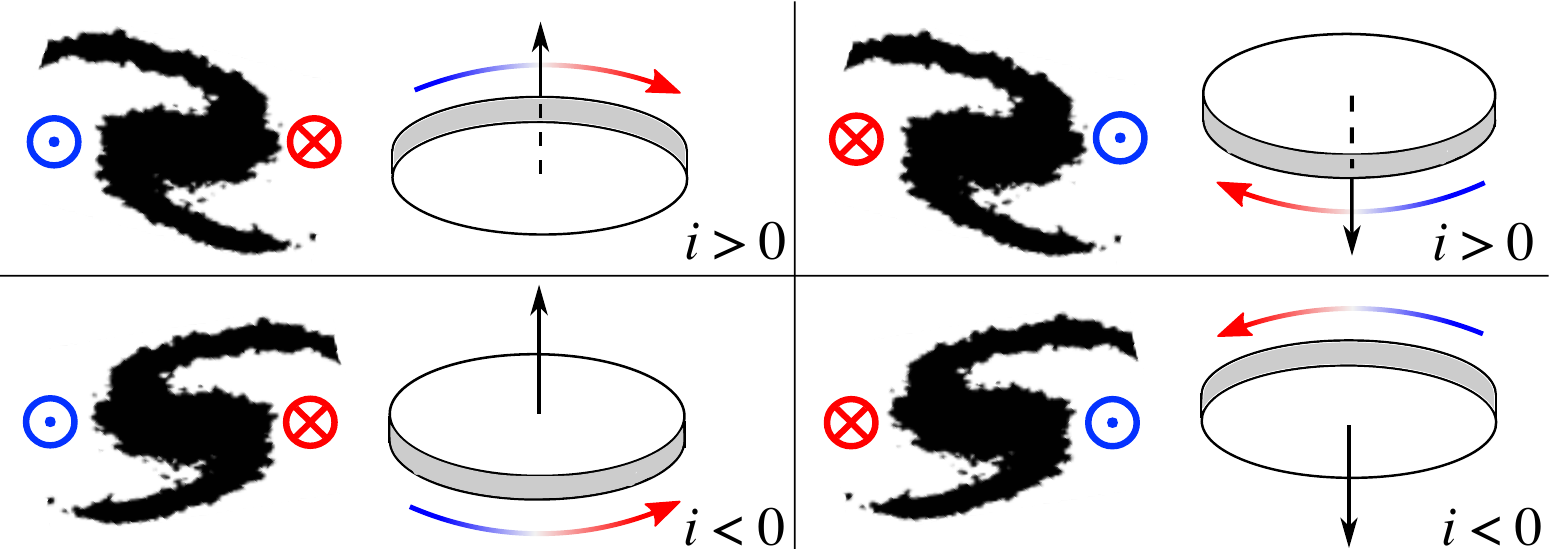}
    \caption{
        Deducing disk tilt from wrapping direction of spiral arms and 
        measured galaxy disk rotation.  
        On each schematic diagram of the spiral arms, 
        the blue dot and the red cross represent the 
        blueshifted and the redshifted sides 
        of the measured disk rotation.  
        The deduced disk tilt is shown on the right.  
        The curved color arrow indicates 
        the rotation direction of each disk, 
        and the straight black arrow shows the 
        corresponding angular momentum vector.
        }
    \label{fig:rc_arm_flip} 
\end{figure}

In this paper, 
we will demonstrate a method to 
independently deduce the disk tilt, 
i.e., the sign of disk inclination, \textit{a priori};  
Figure~\ref{fig:rc_arm_flip} illustrates this concept.  
Because disks rotate differentially, 
and only trailing spirals are long-lived
\citep{Carlberg1985}, 
spiral arms generally trail the direction of rotation.  
Hence, 
using the observed wrapping direction of spiral arms
and the direction of disk rotation
measured from the rotation curve, 
we can determine which way the disk tilts on the sky.

This work examines five galaxy--quasar pairs 
from \citetalias{Ho2017} and \citetalias{Martin2019}.  
We present new, high-resolution images for the five galaxies,
including optical images from the Hubble Space Telescope (HST)
and adaptive-optics-corrected (AO-corrected), 
near infrared images from the Keck Observatory. 
Section~\ref{sec:observations}
briefly describes the imaging observations.  
Section~\ref{sec:disk_tilt_model} shows that
these new images reveal the wrapping direction of the spiral arms 
for four galaxies.  
Together with the rotation curves 
(presented in \citetalias{Ho2017}),
we determine how the disks tilt
and examine whether/how the tilts
constrain the radial inflow model in \citetalias{Ho2017}.  
Section~\ref{sec:implication} discusses 
the implications of our results,
which we summarize in Section~\ref{sec:conclusion}.
Throughout the paper, we adopt the cosmology from \citet{PlanckXIII2015}, 
with $h = 0.6774$, $\Omega_m = 0.3089$, 
$\Omega_\Lambda = 0.6911$, and $\Omega_b = 0.0486$.

\section{Data and Observations}
\label{sec:observations}

We present new imaging observations of five galaxies from
the galaxy--quasar pairs in 
\citetalias{Ho2017} and \citetalias{Martin2019}.  
For these five $z\approx0.2$, blue galaxies,
\citetalias{Ho2017} and \citetalias{Martin2019} 
detected \mgII\ absorption
along the quasar sightlines
and measured the galaxy rotation curves
using the Low Resolution Imaging Spectrometer
(LRIS; \citealt{Oke1995,Rockosi2010}) at Keck
and the Double Imaging Spectrograph
at the Apache Point Observatory 3.5m telescope.\footnote{
    Instrument specifications can be found in the manual
    written by Robert Lupton, which is available at
    \url{http://www.apo.nmsu.edu/35m_operations/35m_manual/Instruments/DIS/DIS_usage.html\#Lupton_Manual}
}
Table~\ref{tb:spec_results} summarizes 
the key spectroscopic measurements of the five systems.  
This paper presents
new optical images from HST Wide Field Camera 3 (WFC3)
and/or AO-corrected, near infrared images
from the Keck NIRC2 camera 
of the galaxies. 
These high resolution images
reveal galaxy structural features
that cannot be recognized in the SDSS images,
e.g., the wrapping of spiral arms,
the presence of companion galaxies, etc.  
Table~\ref{tb:im_obs} lists
the imaging observations of the galaxies.
We describe the HST WFC3 and Keck NIRC2 observations
in Section~\ref{ssec:hst_obs} and \ref{ssec:nirc2_obs},
respectively.
Section~\ref{ssec:environment} briefly
describes the environment of the galaxies.

\begin{center}
\begin{deluxetable*}{llcccccccccc}
\tablecaption{Target Information and Summary of Spectroscopic Measurements}
\tabletypesize{\footnotesize}
\tablewidth{0pt}
\tablehead{
\colhead{Quasar Name} &
\colhead{Galaxy Name} &
\colhead{$i$} &
\colhead{$\alpha$} &
\colhead{$z_{gal}$} &
\colhead{$b$} &
\colhead{$R_g$} &
\colhead{$W^{2796}_r$} &
\colhead{$v_D^{\mathrm{Mg\ II}}$} &
\colhead{$\Delta v_{2796}$} & 
\colhead{$\Delta v_{2796}^{intr}$} &
\colhead{$v_\mathrm{rot}$}
\\
\colhead{} &
\colhead{} &
\colhead{(\deg)} &
\colhead{(\deg)} &
\colhead{} &
\colhead{(kpc)} &
\colhead{(kpc)} &
\colhead{(\AA)} & 
\colhead{(\kmstb)} &
\colhead{(\kmstb)} & 
\colhead{(\kmstb)} &
\colhead{(\kmstb)}
}
\decimalcolnumbers
\startdata
J091954+291408 & J091954+291345 & 73 & 15 & 0.23288 & 88 & 115 & $0.52^{+0.02}_{-0.02}$ & $137\pm4$  & [$-51$, $292$] & [$7$, $235$] & 250\\
                                                    & & & & & &     & $0.22^{+0.01}_{-0.01}$ & $-154\pm9$ & [$-266$, $-8$] & [$-208$, $-66$] & \tablenotemark{a}\\
J102907+421752 & J102907+421737 & 50 & 19 & 0.26238 & 65 & 69 & $0.12^{+0.03}_{-0.03}$ & $-53\pm15$ & [$-134$, $34$] & [$-76$, $-24$] & 155\\
J123049+071036 & J123049+071050 & 38 & 4  & 0.39946 & 98 & 98 & $0.08^{+0.02}_{-0.01}$ & $-91\pm34$\tablenotemark{b} & [$-172$, $-21$] & [$-115$, $-79$] & 190\\
J124601+173156 & J124601+173152 & 63 & 11 & 0.26897 & 19 & 20 & $0.31^{+0.03}_{-0.03}$ & $-299\pm11$ & [$-404$, $-200$] & [$-314$, $-290$] & 60\\
J142501+382100 & J142459+382113 & 61 & 8  & 0.21295 & 83 & 85 & $0.24^{+0.03}_{-0.02}$ & $9\pm7$     & [$-128$, $221$] & [$-37$, $130$] & 190\\
\enddata
\tablenotetext{a}{\small
        We attribute this blue, weak absorption component to a red galaxy within the group.  
        See Paper I or Section~\ref{ssec:environment} for details.
        }
\tablenotetext{b}{\small
        The absorption system falls in a part of the LRIS spectrum without arc lamp lines. 
        The dispersion solution is extrapolated about 200\AA\ beyond the last arc line, 
            introducting potential systematic error.
        }
\tablecomments{
    (1) Name of the quasar.  
    (2) Name of the galaxy.
    (3) Inclination of the galactic disk.
    (4) Azimuthal angle, the angle between the galaxy major axis and the quasar sightline.  
    (5) Galaxy systemic redshift measured from emission lines.
    (6) Sightline impact parameter.
    (7) Galactocentric radius.
    (8) Rest-frame equivalent width of \mgIIdbl.
    (9) \mgII\ Doppler shift measured from line profile fitting.
    (10) Measured velocity range. 
    (11) Intrinsic velocity range 
         (corrected for the line broadening effect due to the instrumental resolution).
    (12) Asymptotic galaxy rotation speed.
}
\label{tb:spec_results}  
\end{deluxetable*}

\end{center}

\begin{center}
\begin{deluxetable*}{llcccccl}
\tablecaption{Imaging Observations\label{tb:im_obs}}
\tabletypesize{\footnotesize}
\tablewidth{0pt}
\tablehead{
\colhead{Target Galaxy} &
\colhead{Instrument} &
\colhead{Exposure Time} &
\colhead{Filter} &
\colhead{Field-of-view} & 
\colhead{Plate Scale} & 
\colhead{PSF FWHM\tablenotemark{a}} &
\colhead{Observing Dates}
\\
\colhead{} &
\colhead{} &
\colhead{(s)} &
\colhead{} &
\colhead{($''$)} &
\colhead{($''$ pixel$^{-1}$)} &
\colhead{($''$)} &
\colhead{}
}
\startdata
J091954+291345 & Keck/NIRC2     & 600  & $K_s$ & $40'' \times 40''$                    & 0.039686 & 0.20 & 2017 Jan 28 \\
J102907+421737 & HST/WFC3 UVIS  & 1716 & F390W & $162'' \times 162''$\tablenotemark{b} & 0.04     & 0.07 & 2017 Jan 25 \\
               & HST/WFC3 UVIS  & 700  & F814W & $162'' \times 162''$\tablenotemark{b} & 0.04     & 0.07 & 2017 Jan 25 \\
               & Keck/NIRC2     & 600  & $K_s$ & $40'' \times 40''$                    & 0.039686 & 0.15 & 2017 Jan 26 \\
J123049+071050 & Keck/NIRC2     & 600  & $K_s$ & $40'' \times 40''$                    & 0.04     & 0.12 & 2015 May 6 \\
J124601+173152 & HST/WFC3 UVIS  & 1629 & F390W & $162'' \times 162''$\tablenotemark{b} & 0.04     & 0.07 & 2017 Jan 28 \\
               & HST/WFC3 UVIS  & 700  & F814W & $162'' \times 162''$\tablenotemark{b} & 0.04     & 0.07 & 2017 Jan 28 \\
               & Keck/NIRC2     & 600  & $K_s$ & $40'' \times 40''$                    & 0.039686 & 0.23 & 2017 Jan 26 \\ 
               & Keck/NIRC2     & 1200 & $K_s$ & $40'' \times 40''$                    & 0.039686 & 0.31 & 2017 Apr 13 \\
J142459+382113 & HST/WFC3 UVIS  & 1674 & F390W & $162'' \times 162''$\tablenotemark{b} & 0.04     & 0.07 & 2017 Jun 5 \\
               & HST/WFC3 UVIS  & 700  & F814W & $162'' \times 162''$\tablenotemark{b} & 0.04     & 0.07 & 2017 Jun 5 \\
               & Keck/NIRC2     & 600  & $K_s$ & $40'' \times 40''$                    & 0.039686 & 0.14 & 2015 May 6 \\
\enddata
\tablenotetext{a}{\small The full-width-half-maximum (FWHM) of the point spread function (PSF).}
\tablenotetext{b}{\small The WFC3/UVIS channel has a rhomboidal field-of-view.}
\end{deluxetable*}
\end{center}

\subsection{HST WFC3 Observations}
\label{ssec:hst_obs}

We imaged three galaxies 
using the WFC3/UVIS channel
and the F390W and F814W broadband filters
(Cycle 24, PID: 14754, PI: C.~L.~Martin).  
We list the exposure time for each galaxy in
Table~\ref{tb:im_obs}.  
The table also includes the field-of-view, the plate scale,
and the full-width-half-maximum (FWHM) of the point spread function (PSF)
of the UVIS channel.

Individual data frames were retrieved from 
MAST\footnote{\url{http://archive.stsci.edu}}.
These data frames were already reduced and calibrated 
by the standard WFC3 calibration pipeline \texttt{calwf3}.\footnote{
    Documentation of \texttt{calwf3} can be found at 
    Gennaro, M., et al. 2018, “WFC3 Data Handbook”, Version 4.0, 
    (Baltimore: STScI).} 
For the F390W data frames, 
we drizzled the three dithered images of each target using
\texttt{DrizzlePac} \citep{Fruchter2010}.\footnote{
    Documentation of \texttt{DrizzlePac} can be found at 
    \citet{Gonzaga2012}.
    }
But for the F814W data frames, 
because we only took two exposures per target,
cosmic ray removal during drizzling was sub-optimal.
Therefore, we used L.A.Cosmic \citep{vanDokkum2001}
to remove the cosmic rays from individual F814W frames.
Then we drizzled the cleaned frames to
produce the science images for individual targets.

\subsection{Keck NIRC2 Observations}
\label{ssec:nirc2_obs}

We observed all five galaxies with the NIRC2 camera
on the Keck II telescope, 
using the paired quasar as the tip-tilt reference 
for the Keck laser guide star adaptive optics system 
(LGSAO; \citealt{vanDam2006,Wizinowich2006}).  
\citetalias{Martin2019} described 
the observation and the data reduction in detail.
In brief,
we observed each galaxy using the $K_s$ broadband filter,
which centered at 2.146 \micron;
Table~\ref{tb:im_obs} lists the $K_s$ observation. 
We reduced the images using the data reduction pipeline
provided by the
UCLA/Galactic center group \citep{Ghez2008}.  
The final images were
dark corrected, flat-fielded, sky subtracted,
and corrected for geometrical distortion.

\subsection{Galaxy Environment}
\label{ssec:environment}

Galaxy group environment affects the 
properties of \mgII\ absorption,
such as the absorption strength and velocity dispersion
\citep{Bordoloi2011,Johnson2015,Nielsen2018},
and possibly leads to uncertain host galaxy assignment 
of the \mgII\ absorption.   
We have flagged three of the five target galaxies
as potential group members \citepalias{Ho2017,Martin2019}.

J124601+173152 is in a rich environment,
with three brighter galaxies having consistent photometric redshifts.
While our target galaxy is the closest to the quasar sightline,
both \citetalias{Ho2017} and \citetalias{Martin2019}
flagged this system because
the host identification for the \mgII\ absorption may not be unique.

Because we lack spectroscopic redshifts for 
all but the closeset galaxies in our fields, 
we cannot classify the environment of the absorbers 
as \citet{Chen2009} and \citet{Johnson2013} did.
Nevertheless, 
we searched for potential group members 
using the SDSS images and photometric redshifts 
of the galaxies in each field,
and we found two other target galaxies potentially in groups.  
We found there is a potential group member 
near the target galaxy J102907+421737.  
Because our target is three times closer
to the sightline than the other bright galaxy in the field,
the association of \mgII\ absorption in the 
J102907+421752 sightline with our target galaxy is secure.  
Another target galaxy, J091954+291345, 
potentially forms a group with two red galaxies
$4\farcs4$ and $8\farcs7$ away,
both of which have photometric redshifts consistent with our target. 
The spectrum of our paired quasar detects two absorption systems.
Following \citetalias{Ho2017} and \citetalias{Martin2019},
we assign the stronger system to the blue galaxy (our target), 
and we consider the weaker system as 
potentially not produced by our target.

\section{Individual Galaxies: 
Disk tilt and Gas Kinematic Modeling}
\label{sec:disk_tilt_model}

The new, high-resolution images
reveal the spiral structures for four of the five galaxies.
We deduce how these disks tilt, 
and then we explore how their tilts 
constrain the gas kinematic modeling,
which aims to explain the 
broad \mgII\ velocity range measured.
Section~\ref{ssec:disk_tilt} shows the new images 
and discusses the image processing steps.  
Section~\ref{ssec:model_short} reviews
the key components of the radial inflow model 
from \citetalias{Ho2017}.
Then in Section~\ref{ssec:diskgal_model},
we discuss individual galaxies
regarding how each disk tilts,
and we explore how it combines with the inflow model 
in \citetalias{Ho2017}
and affects the modeled inflow properties.

\subsection{Deducing the Disk Tilt From 
High-Resolution Images}
\label{ssec:disk_tilt}

\begin{sidewaysfigure*}[htb]
    \vspace{-9.5cm}  
    \centering
    \includegraphics[angle=0,width=0.96\textwidth]{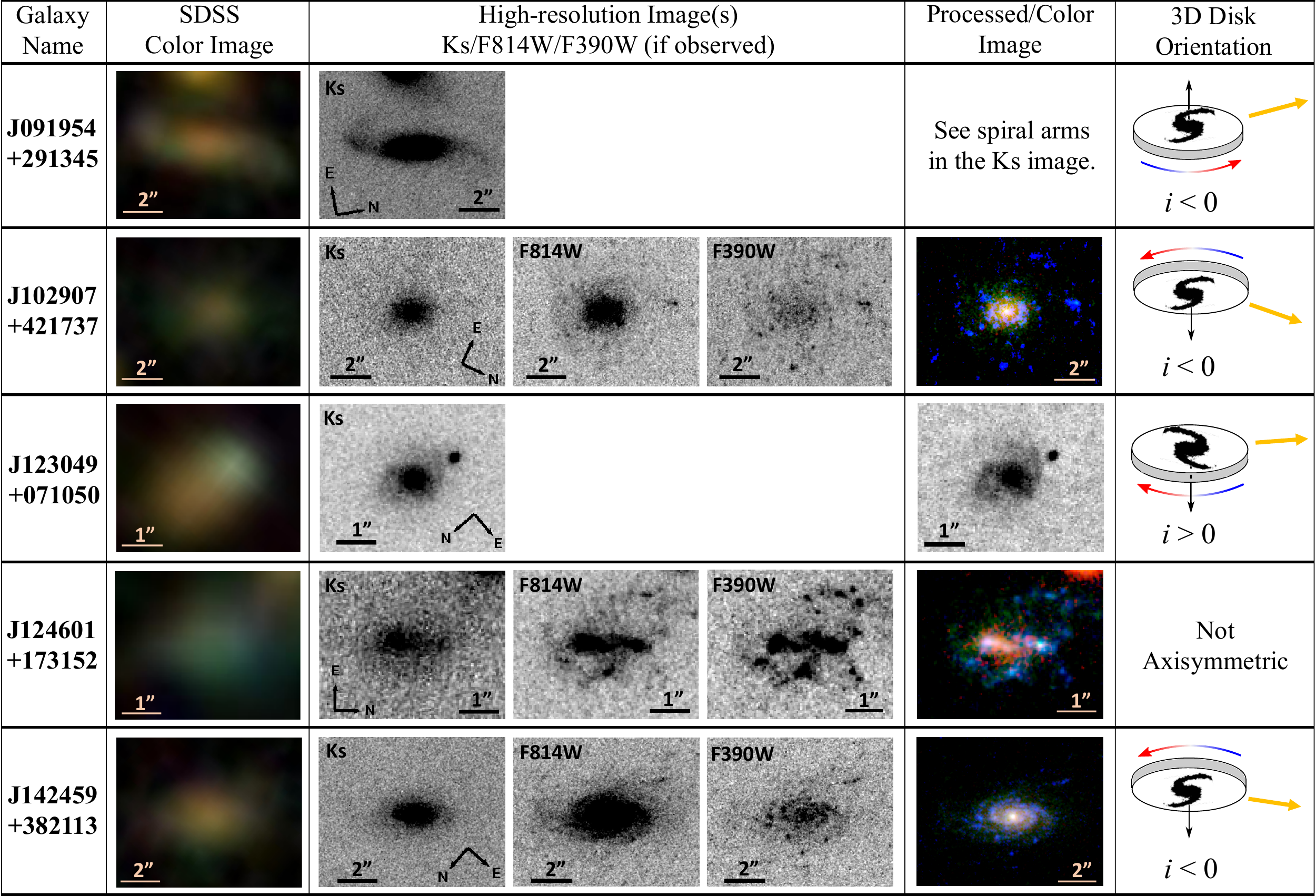}
    \caption{
        High-resolution galaxy images 
        and the deduced 3D disk orientations.
        Each row corresponds to one galaxy,
        and column 1 shows the galaxy name.
        Column 2 demonstrates that 
        the low-resolution SDSS images 
        cannot resolve the galaxy structural features.
        Column 3 shows the high-resolution 
        NIRC2 $K_s$, WFC3 F814W, and WFC3 F390W 
        (if observed) for individual galaxies.
        For galaxies with spiral arms or structures 
        that can hardly be distinguished in the original images,
        column 4 shows the processed images 
        that visually enhanced the structural features.
        In particular, column 3 shows 
        the three-color images for galaxies
        observed in all three bands.
        Column 5 illustrates the wrapping direction of the spiral arms,
        the measured disk rotation (the curved arrow), 
        and the deduced 3D disk orientation.
        Each orange arrow points toward the 
        direction of the quasar sightline.
        The images for individual galaxy have the same scale
        and the same orientation; 
        each galaxy major axis aligns with the image x-axis,
        and the quasar sightline lies in the positive x-direction.
        }
    \label{fig:galim}
\end{sidewaysfigure*}

Figure~\ref{fig:galim} shows the images of individual galaxies
and the deduced 3D disk orientations.  
We rotate the galaxy images such that
each galaxy major axis aligns with the image x-axis,
and the quasar sightline lies in the positive x-direction.
Each schematic diagram in the last column
illustrates the wrapping direction of spiral arms
and direction of the disk rotation (curved, blue/red arrow).
Following the concept illustrated by Figure~\ref{fig:rc_arm_flip}
and the definition in \citetalias{Martin2019},
we deduce the sign of disk inclination,
i.e., which way the disk tilts.

Contrasting our high-resolution images 
with the color images from 
the Sloan Digital Sky Survey (SDSS) in Figure~\ref{fig:galim}
highlights the resolving power of our new images.
For the SDSS images,
the median full-width-half-maximum (FWHM)
of the point spread function (PSF)
in the $r$-band is $1\farcs3$.  
This is an order of magnitude higher than that 
for both WFC3 optical and the NIRC2 $K_s$ images.  
Therefore, these images accentuate 
the capability of the high-resolution imaging to 
reveal structures and companion galaxies
that low-resolution imaging fails to resolve.

\begin{figure}[htb]
    \centering
    \includegraphics[width=1.0\linewidth]{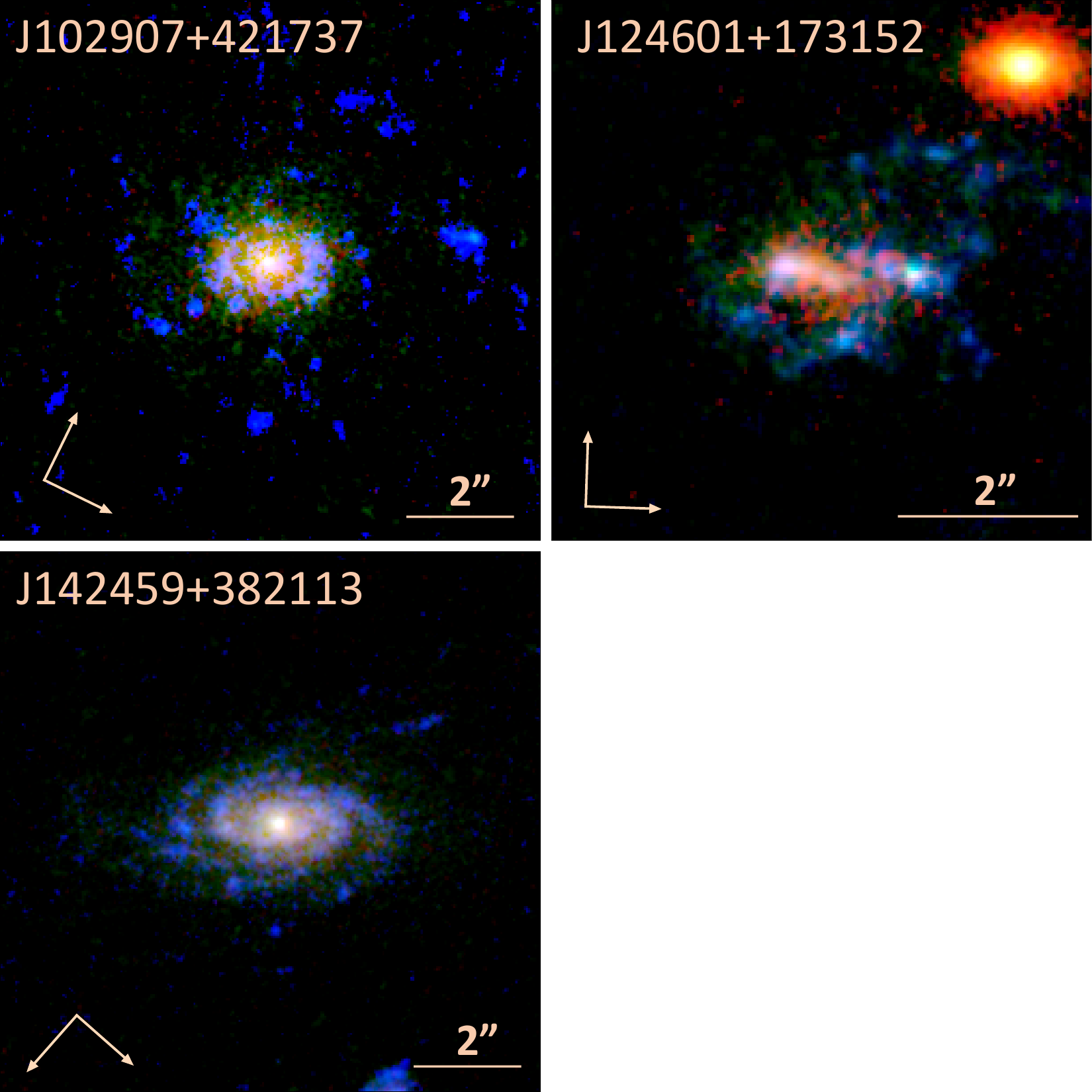}
    \caption{Composite F390W (blue), F814W (green), 
                and $K_s$ (red) images.
                Both J102907+421737 and J142459+382113 
                show spiral arms and bluer outer disks. 
                J124601+173152 shows two bright clumps along
                the N-S direction.
        }
    \label{fig:colorim} 
\end{figure}

When the spiral arms 
are not prominent in the high-resolution images,
we apply additional image processing steps
to visually enhance the structural features.
We create three-color images for
J102907+421737, J124601+173152, and J142459+382113,
all of which are observed in
F390W, F814W, and $K_s$ bands
(Figure~\ref{fig:colorim}).
We use blue, green, and red colors to represent 
the three bands respectively.\footnote{
    We register the images before creating the color images.
    We resample the pixels to the pixel scale of the WFC3 images
    and align all images using 
    reference point sources in the common field.
    }
The color images also allow us to visualize the 
the spatial distribution 
of different stellar populations.  
In addition, 
the F390W image of J102907+421737
and the $K_s$ image of J123049+071050
only show weak traces of spiral arms.
So, we process the images as follows
to enhance the spiral features.
First, 
we use GALFIT \citep{Peng2002} to
model the smooth galaxy emission component with
an exponential disk surface brightness (SB) profile
and convolve the model
with the instrumental PSF.
We model the instrumental PSF of the HST WFC3
using Tiny Tim \citep{Krist2011},
and for the $K_s$ image,
we fit a Moffat profile to our paired quasar.
The residual image,
i.e., the difference between the original image and
the convolved galaxy SB model,
shows traces of spiral arms.
Then, 
we apply a median filter to the residual
and add it to the original F390W or $K_s$ images.  
For J102907+421737,
we use the enhanced F390W image to 
create the three-color image.
We present the processed images 
under the ``Processed/Color Image'' column 
in Figure~\ref{fig:galim}.

\subsection{Brief Description of the Disk Model 
and Radial Inflow}
\label{ssec:model_short}

The simple inflow model in \citetalias{Ho2017} 
combines
the tangential (rotational) and radial (inflow) velocity components 
on the disk plane.  
For the radial inflow component, 
we adopt a constant radial velocity,
\begin{equation}
    \boldsymbol{v}_R(R,z) = v_R \boldsymbol{\hat{R}}
    ~,
    \label{eq:vr}
\end{equation}
where $v_R < 0$ for inflow. 
We model the rotation component as 
\begin{equation}
    \boldsymbol{v}_\phi(z) = v_{\mathrm{rot}}  \exp(-|z|/h_v) \boldsymbol{\hat{\phi}}
        ~,
    \label{eq:sk-rotation}
\end{equation}
where $v_{\mathrm{rot}}$ represents 
the asymptotic galaxy rotation speed measured,
and the velocity scale height $h_v$ 
introduces a vertical velocity gradient 
to create a rotational lag  
above and below the disk midplane.  
\citetalias{Ho2017} discussed two cases:
(1) $h_v \rightarrow \infty$, i.e., without a velocity gradient,
and (2) a fiducial $h_v$ value of 10 kpc,
which produced vertical velocity gradients of
11 to 26 \kms\ per kpc for the galaxy sample.
Because the fiducial 10 kpc value creates
velocity gradients consistent with measurements 
of extraplanar gas, 
we follow \citetalias{Ho2017} 
and adopt the same fiducial \hv\ value
when we consider disks with lagging rotation.
The velocity gradients have been measured 
to about 10 kpc vertically and 
over 20 kpc radially in nearby galaxies
\citep{Benjamin2002,Oosterloo2007,Marasco2011,
Zschaechner2011,Zschaechner2012,Gentile2013,Kamphuis2013}, 
so our approach extrapolates into a regime unconstrained by 
direct measurements.

We obtain the LOS velocity $v_\mathrm{los}$ 
by taking the dot product between each of 
Equations~(\ref{eq:vr}) and (\ref{eq:sk-rotation}) 
with the vector describing the quasar sightline.
The impact parameter $b$ and azimuthal angle $\alpha$ 
set the orientation of the sightline 
with respect to the galaxy disk.  
The sign of the disk inclination $i$
describes which way the disk tilts,
whereas the magnitude of the disk inclination and 
the modeled thickness \Heff\ (measured from the disk midplane)
limit the pathlength along the sightline that intersects the disk.  
Along this sightline path,
the bluest and the reddest ends of the $v_\mathrm{los}$ 
define the resultant LOS velocity range.

\subsection{Disk Tilt of Individual Galaxies and their 
Gas Kinematic Modeling}
\label{ssec:diskgal_model}

Using new individual galaxy images
and the measured galaxy rotation curves 
(presented in \citetalias{Ho2017}),
we determine which way the disk tilts.
Then, we investigate whether and/or how 
the deduced disk tilt
constrain the gas kinematic modeling and
the inflow properties.  
Specifically,
we explore whether the rotating disk model,
without and with radial inflow,
can reproduce the broad \mgII\ LOS velocity range 
measured along the quasar sightlines.

For individual galaxies,
we show position--velocity (PV) diagrams,
i.e., $D_\mathrm{los}$ vs.~$v_\mathrm{los}$,
for the rotating-disk-only model (no radial inflow)
and models with radial inflow for disks at
$i>0$ and $i<0$.  
Contrasting the two tilts demonstrates 
how the \textit{a priori} knowledge of the disk tilt
affects the modeling.  
In each PV diagram, 
$D_\mathrm{los} = 0$ represents the disk midplane,
and $D_\mathrm{los} < 0$ ($> 0$) represents
the sightline at the near (far) side of the disk.
The cyan shaded region represents the 
measured LOS velocity range of the \mgII\ absorption.  
The cyan hashes indicate the uncertainties of the 
measured velocity range at the blue and red ends.
The gray shaded region that enclosed 
the modeled $v_\mathrm{los}$ (black curve)  
shows the line broadening effect that matches 
the LRIS spectral resolution.
The modeled LOS velocity range (gray shaded region)
agrees with the measurement only if 
both the bluest and reddest end of the modeled $v_\mathrm{los}$ 
fall within the uncertainties 
of the two ends of the measurement (cyan hashes).  
To illustrate the known problem 
that thin disks cannot explain broad velocity ranges, 
the two yellow markers in the PV diagram (left) indicate
where the sightline
intersects the boundary of an \Heff\ = 1 kpc disk.  
Along the short sightline path that intersects the thin disk,
because of the small variation in the LOS velocity, 
a thin disk can only produce a narrow velocity range.

\subsubsection{J091954+291345}
\label{sssec:j0919_im}

\begin{figure*}[htb]
    \centering
    \includegraphics[width=1.0\linewidth]{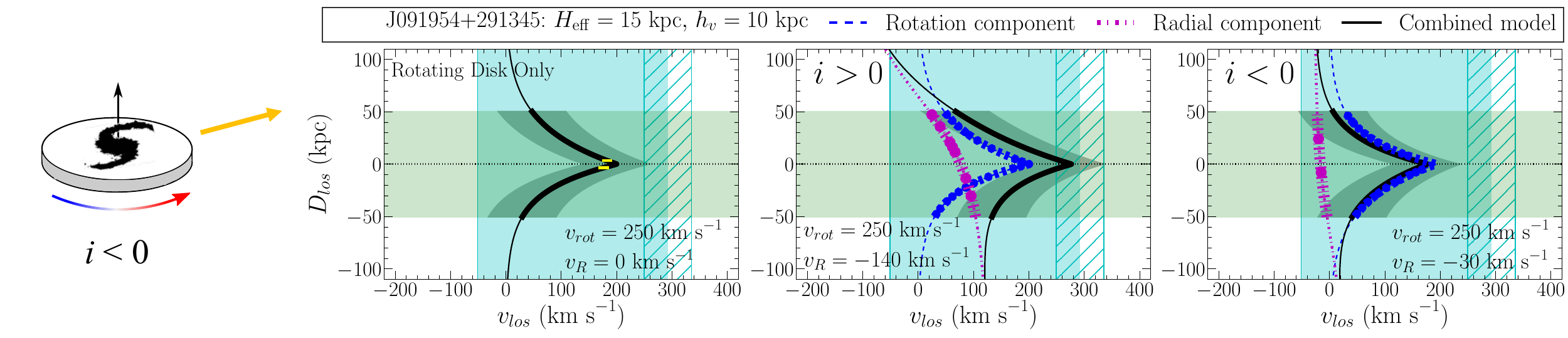}
    \caption{
        J091954+291345: Disk tilt  
        tightens constraint on radial inflow speed. 
        The galaxy disk is negatively inclined
        (the schematic diagram and Section~\ref{sssec:j0919_im}).
        A rotating disk of \Heff\ = 15 kpc and \hv\ = 10 kpc 
        can reproduce the LOS velocity range 
        of the absorption (left), 
        and a thinner disk cannot reproduce the blue absorption end.
        For example, the pathlength within 
        an \Heff\ = 1 kpc disk is short, producing only
        a narrow velocity range (yellow markers).  
        Disks with positive (middle) and negative (right) 
        inclinations
        allow maximum radial inflow speeds of 140\kms\ and 30\kms\ 
        respectively.  
        Otherwise, the absorption becomes too red or not red enough
        at the disk midplane.  
        Hence, deducing the disk as negatively inclined limits 
        the inflow speed to 30\kms.        
        }
    \label{fig:j0919_model} 
\end{figure*}

The galaxy J091954+291345 clearly shows the spiral arms
in the $K_s$ image.
Together with the direction of disk rotation
determined from the rotation curve,
we deduce that the disk has a negative inclination.

A disk without a vertical velocity gradient 
can explain the over 300\kms\ LOS velocity range, 
but that would require
a 500-kpc thick ``disk'' (\Heff\ = 250 kpc)
to match the blue absorption end.  
With \hv = 10 kpc, a
significantly thinner disk of \Heff\ = 15 kpc
(left panel of Figure~\ref{fig:j0919_model})
can explain the same velocity range.  
Further reducing \Heff\ would make the absorption
not blue enough to match the measurement.  
Considering also radial inflow for this \Heff\ = 15 kpc disk,
a positively inclined disk would allow an inflow speed of 
140\kms\ before the absorption becomes too red 
at the disk midplane, 
but the disk would need to be thicker to 
match the blue absorption end (middle panel).
On the other hand, 
a negatively inclined disk can reproduce 
the measured velocity range with radial inflow of
no more than 30\kms\ (right panel);
otherwise the absorption would not 
be red enough at the disk midplane.
Because the two disk tilts permit significantly different 
radial inflow scenarios,
the deduced negative disk inclination tightens the 
constraint on the radial inflow speed to at most 30\kms.

\subsubsection{J102907+421737}
\label{sssec:j1029_im}

For J102907+421737,
the F390W image shows the spiral arms most clearly,
and we visually enhance 
the spiral features and create the color image
as described in Section~\ref{ssec:disk_tilt}.  
The color image also shows that the 
outer disk is bluer than the inner disk.  
This suggests that the outer disk is younger 
than the inner disk
\citep{deJong1996,Bell2000,MacArthur2004},
indicating an inside-out disk growth.
From the revealed wrapping direction 
of the spiral arms and the measured disk rotation, 
we deduce that the disk has a negative inclination.

\begin{figure*}[htb]
    \centering
    \includegraphics[width=1.0\linewidth]{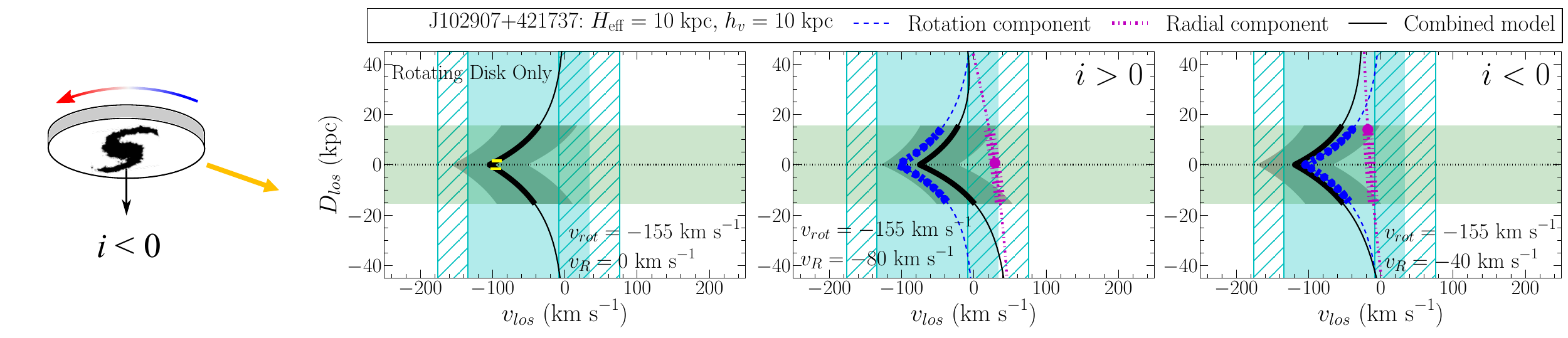}
    \caption{
        J102907+421737: Disk tilt  
        tightens constraint on radial inflow speed.  
        The galaxy has a negatively inclined disk
        (the schematic diagram and Section~\ref{sssec:j1029_im}).
        A rotating disk (left) of \hv\ = 10 kpc with 
        \Heff\ = 10 kpc can reproduce 
        the observed LOS velocity range 
        within measurement uncertainties.  
        A sightline that intersects a thinner disk, 
        e.g., an \Heff = 1 kpc disk (yellow markers),
        cannot produce a broad enough velocity range.
        Disks with 
        positive (middle) and negative (right) inclinations 
        allow maximum radial inflow speeds 
        of 80\kms\ and 40\kms\ respectively.  
        Otherwise, the absorption becomes 
        not blue enough or too blue
        at the disk midplane.  
        With the inferred disk inclination being negative,
        the radial inflow speed cannot exceed 40\kms.
        }
    \label{fig:j1029_model} 
\end{figure*}

A disk without a velocity gradient
can explain the $\sim$100\kms\ measured LOS velocity range.
However, the disk would need an \Heff\ of 65 kpc 
to bring the absorption close to the galaxy systemic velocity
in order to match the red absorption end.
With \hv\ = 10 kpc,
a thinner rotating disk of \Heff\ = 10 kpc 
can reproduce the LOS velocity range of the absorption
within measurement uncertainties
(left panel of Figure~\ref{fig:j1029_model}).
Further reducing the disk thickness would make 
the absorption not red enough to match with the measurement.
If we also consider radial inflow on the 
\Heff\ = 10 kpc disk,
then the inflow speed
for a positively (negatively) inclined disk
cannot exceed 80 (40) \kms.
Otherwise, the absorption becomes not blue enough (too blue)
at the disk midplane; see the middle (right) panel.  
Hence, the inferred negative disk inclination limits 
the radial inflow speed to at most 40\kms.

\subsubsection{J123049+071050}
\label{sssec:j1230_im}

The high-resolution $K_s$ image resolves 
the target ``galaxy'' into two objects: 
a galaxy with spiral arms and a possible satellite.
The galaxy also has weak traces of spiral arms,
and we visually enhance the arms 
as described in Section~\ref{ssec:disk_tilt}.
From the wrapping direction of the spiral arms 
and the direction of disk rotation,
we deduce that the disk has a positive inclination.

A rotating disk without a vertical velocity gradient
can reproduce the 160\kms\ measured velocity range,
but the ``disk'' has to be 260-kpc thick (\Heff\ = 130 kpc)
to bring the red absorption end close to the 
galaxy systemic velocity.  
In contrast, the fiducial model with a vertical velocity gradient
only requires a 10-kpc thick disk (\Heff\ = 5 kpc)
to reproduce the velocity range spanned by the absorption.

\begin{figure*}[htb]
    \centering
    \includegraphics[width=1.0\linewidth]{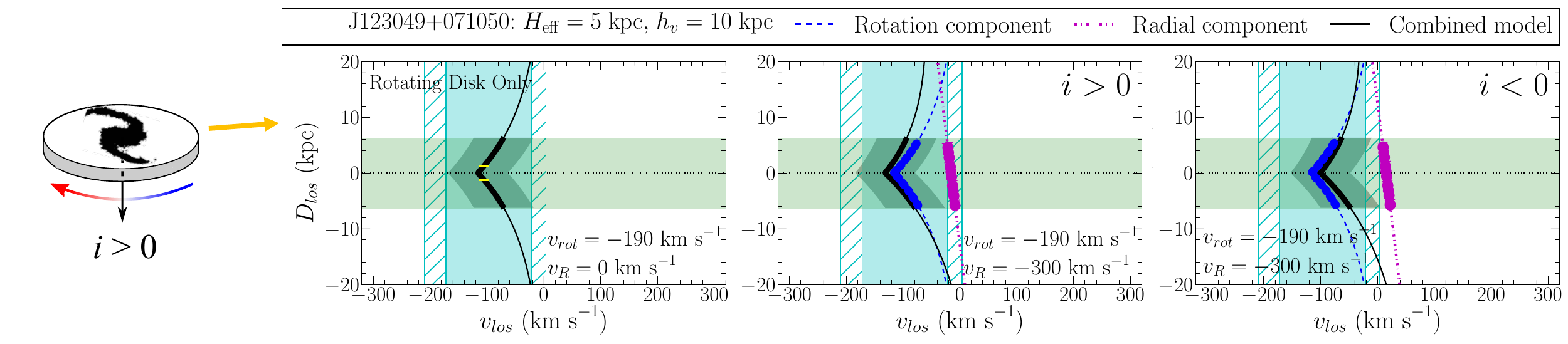}
    \caption{
        J123049+071050: Disk tilt cannot tighten constraint 
        on radial inflow speed.  
        We deduce the disk as being positively inclined 
        (the schematic diagram and Section~\ref{sssec:j1230_im}).
        A rotating disk (left) of \Heff\ = 5 kpc and \hv\ = 10 kpc 
        can explain the velocity range of the absorption,
        and a thinner disk cannot create broad enough absorption
        with the red absorption end close to 
        the galaxy systemic velocity.
        Alternatively,
        regardless of which way the disk tilts,
        a radial inflow of 300\kms\
        changes the velocities at both the blue and the red ends
        by no more than 20\kms\ (middle and right).
        Both tilts can still reproduce the measured 
        LOS velocity range within measurement uncertainties.
        Hence, knowing the disk tilt \textit{a priori} cannot tighten
        the constraint on the radial inflow speed.
        }
    \label{fig:j1230_model} 
\end{figure*}

Knowing the disk tilt \textit{a priori} 
cannot rule out radial inflow 
nor constrain the radial inflow speed
for this galaxy.  
Even as the inflow speed increases from zero to 300\kms, 
the LOS velocities at both the bluest and the reddest ends change 
by no more than 20\kms.  
This 20\kms\ shift is allowed by 
the measurement uncertainties. 
This means that regardless of which way the disk tilts,
an inflow speed as large as  300\kms\ can 
always explain the measured velocity range. 
The reason of why the LOS velocity is insensitive
to both the radial inflow speed and  the disk tilt
is because of the low disk inclination ($i = 38$\deg)
and small azimuthal angle ($\alpha=4$\deg).  
Section~\ref{sec:implication} discusses
how  both factors affect 
the effectiveness of constraining the radial radial speed.

\subsubsection{J124601+173152}
\label{sssec:j1246_im}

Both F390W and F814W images of J124601+173152
reveal that the galaxy has an irregular, clumpy structure.
In both F390W and F814W images, 
the galaxy shows two bright clumps along the N-S direction.
However, the $K_s$ image only detects
a single nucleus that lies at the location of the southern clump.
The color image also shows that 
the emission in $K_s$ overlaps with the southern clump only.

Altogether, our images suggest that this galaxy 
has a complicated, clumpy structure 
instead of a typical, axisymmetric disk morphology.
Hence, we do not define how this galaxy ``disk'' tilts,
and we do not create disk models to explain 
the \mgII\ absorption.

\subsubsection{J142459+382113}
\label{sssec:j1424_im}

\begin{figure*}[htb]
    \centering
    \includegraphics[width=1.0\linewidth]{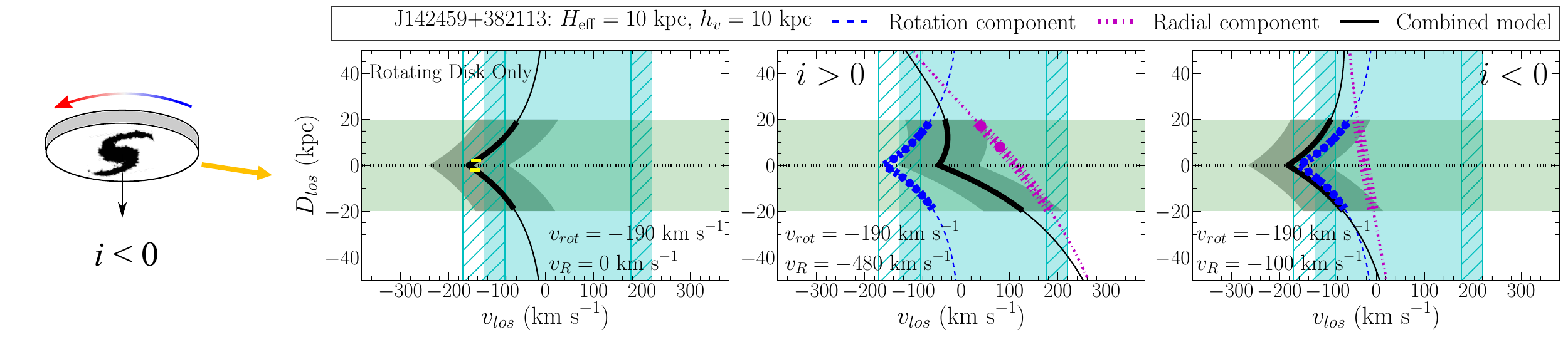}
    \caption{
        J142459+382113: Disk tilt excludes radial inflow detection.
        The disk is negatively inclined
        (the schematic diagram and Section~\ref{sssec:j1424_im}).
        Regardless of the thickness of the disk,
        disk rotation alone (left) cannot reproduce 
        the broad absorption that spans both sides 
        of the galaxy systemic velocity,
        and the rotation produces bluer absorption
        near the disk midplane than measured.
        Only a positively inclined disk (middle) 
        can reproduce the LOS velocity range.
        However, the disk is negatively inclined.
        Introducing radial inflow 
        produces even bluer net absorption (right) 
        than a rotating disk (left) near the midplane.
        We demonstrate this problem using
        a fiducial inflow speed of 100\kms.    
        Hence, the negatively inclined disk of this galaxy 
        excludes the detection of radial inflow in the sightline.
    }
    \label{fig:j1424_model} 
\end{figure*}

Both F390W and F814W images of J142459+382113
unambiguously detect the spiral arms.
From the wrapping direction of the arms
and the measured disk rotation,
we deduce the disk as being negatively inclined.

Moreover, the color image clearly shows that
the outer disk is bluer than the inner disk.
The presence of the color gradient is also supported by the
larger disk scalelengths measured at the bluer bands.
The color gradient indicates that
the outer disk has a younger stellar population
and a lower metallicity than the inner disk, 
suggesting the disk grows inside-out.  
In Section~\ref{sec:implication},
we will discuss the implications
of this disk growing inside-out
together with the result from our inflow modeling.

The \mgII\ absorption detected in the quasar sightline 
has a small Doppler shift.
Since the broad velocity range spans both sides 
of the galaxy systemic velocity, 
neither the line broadening by instrumental resolution
nor any structure with only rotation 
can explain the two-sided absorption.  
A rotating disk also produces bluer absorption than measured
near the disk midplane.
The left panel of Figure~\ref{fig:j1424_model} illustrates 
this problem using 
a disk with a vertical velocity gradient.  
A disk without a velocity gradient makes the problem worse,
because the absence of the gradient makes 
the projected rotation velocity
peak at a higher velocity (i.e., bluer) along the LOS.
Therefore, 
while the radial inflow model 
can produce two-sided absorption
that a rotation-only structure fails to produce,
solving the problem of excess blue absorption
from the disk rotation
requires the LOS velocities 
from the inflow and the rotation components to have
opposite signs near the midplane.

Only a positively inclined disk 
can achieve this configuration.  
The middle panel of Figure~\ref{fig:j1424_model} 
illustrates that the radial inflow (rotation) component creates
a positive (negative) LOS velocity near the disk midplane,
and reproducing the measured velocity range
requires an \Heff\ = 10 kpc disk
with an exceptionally high radial inflow speed of 480\kms.
The problem is that we deduce the disk as being 
negatively inclined.  
Radial inflow makes the absorption even bluer
than disk rotation alone, 
because both velocity components 
have the same Doppler sign near the midplane.
The right panel illustrates 
this problem using a fiducial inflow speed of 100\kms.
Therefore, for this galaxy,
its negatively inclined disk excludes
the detection of radial inflow in the sightline.

\section{Discussion and Implications}
\label{sec:implication}

Which way a galaxy disk tilts on the sky 
can be deduced from 
the wrapping direction of the spiral arms
and the direction of disk rotation
(Figures~\ref{fig:rc_arm_flip} and \ref{fig:galim}).  
Using a radial inflow model 
with gas spiraling inward on the disk plane
(as described in \citetalias{Ho2017}),
the four examples in Section~\ref{sec:disk_tilt_model}
have demonstrated incorporating the disk tilt
as an extra parameter in modeling the 
circumgalactic gas kinematics.
With \textit{a priori} knowledge on the disk tilt, 
our examples show that
we can constrain the radial inflow speed 
for a galaxy
or even exclude radial inflow.

\subsection{The Projection of Radial Inflow is Not Symmetric About a Disk Flip}
\label{ssec:why_asymmetry}

In the radial inflow model, 
which way the disk tilts
alters the projected LOS velocity 
from an individual velocity component
and results in different LOS velocity ranges.  
The key to this asymmetry is the location where
the sightline intersects the disk tangentially.\footnote{
    The sightline intersects the disk tangentially 
    only at the point where 
    the disk radius equals to the sightline impact parameter $b$.
    In general, at different $D_{\mathrm{los}}$, 
    the sightline intersects the disk at 
    different disk radii larger than $b$.  
    }
At this tangent point, 
the sightline is normal to the radial velocity vector,
producing zero LOS velocity.  
If the disk rotation does not have a vertical velocity gradient,
then this tangent point also produces the 
maximum rotational velocity projection of $v_\mathrm{rot}\sin |i|$.
When the disk flips,
the tangent point switches
from one side to the other side of the disk.\footnote{
    Whether the tangent point is at the near side 
    or the far side of the disk
    does not only depend on which way the disk tilts,
    but also depend on the position of the quasar sightline
    with respect to the galaxy disk.
    }
This affects whether or not the LOS velocities of the rotation
and the inflow components
have the same sign near the disk midplane,
which determines whether   
the radial inflow boosts or cancels the 
projected rotation velocity.
Consequently, flipping the disk alters the 
resultant LOS velocity  
produced by the two velocity components.

J142459+382113 most clearly demonstrates 
how this property affects the modeling.  
As explained in Section~\ref{sssec:j1424_im},
the radial inflow has to cancel the excess blue absorption
produced by the disk rotation.
To achieve this configuration, 
this galaxy disk must have a positive inclination;
the tangent point is at the far side of the disk
($D_\mathrm{los} > 0$), 
so that near the disk midplane,
the rotation and radial inflow components
produce blue and red Doppler velocities, respectively.
However, the disk is negatively inclined,
and hence, this rules out 
the radial inflow detection in this sightline.

\subsection{Constraining Radial Inflow Speed 
or Predicting the Disk Tilt}
\label{ssec:use_asymmetry}

When we use the model to explain the measured LOS velocity range,
the observationally deduced disk tilt allows us to 
constrain the maximum radial inflow speed 
or exclude radial inflow detection.  
For example, the observed disk tilt of 
J091954+291345 (Figure~\ref{fig:j0919_model}) and 
J102907+421737 (Figure~\ref{fig:j1029_model})
constrains the inflow speeds to 30-40\kms, 
instead of $\approx100$\kms\ allowed by the opposite tilt.  
We also exclude radial inflow detection for 
J142459+382113 (Figure~\ref{fig:j1424_model}), 
because its disk tilts in a direction opposite
to that allowed by the model.
Therefore, in principle,
we can use the sign of disk inclination 
as an extra geometrical parameter
while modeling the \mgII\ velocity range along the LOS,
and hence, 
the asymmetry between the disk tilt 
and LOS velocity measurements
helps model the circumgalactic gas kinematics.

However, 
our example of J123049+071050 has revealed a caveat: 
not all resultant LOS velocity ranges
are sensitive to the radial inflow speed
and/or
change significantly when we flip the disk
(Figure~\ref{fig:j1230_model}).  
In fact, this example has demonstrated
that both the magnitude of the
galaxy disk inclination $|i|$ and 
the azimuthal angle $\alpha$ of the quasar sightline 
affect whether or not 
the disk tilt can constrain the radial inflow speed.
We consider these two factors separately as follows.

First, 
since both the rotation and radial inflow velocity vectors
lie on the disk plane,
a less inclined disk produces 
a smaller LOS velocity projection.
For example, for a face-on ($i=0$\deg) disk, 
because both the tangential (rotation) 
and radial (inflow) velocity vectors
are perpendicular to the sightline, 
we always detect zero velocity along the LOS.  
Therefore, 
due to the projection effect, 
the LOS velocity and the velocity range
of a low-inclination disk  
are insensitive to the change in the radial inflow speed.

Second, when the $\alpha$ is small,
two disks of the same inclination $|i|$ 
but with opposite tilts produce similar 
projected LOS velocity from the radial component.  
The projected radial velocity switches sign
where the sightline intersects the disk tangentially, 
and this tangent point switches sides when the disk flips
(Section~\ref{ssec:why_asymmetry}).
But for an $\alpha=0$\deg\ sightline,
the tangent point is on the disk midplane,
hence flipping the disk does not change 
the projected radial velocity along the LOS.
As $\alpha$ increases, 
the tangent point moves away from the disk midplane.  
Flipping the disk then changes 
how the projected radial velocity varies along the LOS,
resulting in different 
resultant LOS velocities and velocity ranges
when we combine the radial and rotation velocity components.

In short, 
a less inclined disk makes the LOS velocity insensitive to 
the change in the radial inflow speed,
and a small azimuthal angle leads to a small difference in the
resultant projected LOS velocity between the two disk tilts.  
Consequently, 
to use the disk tilt to tighten the 
constraint on the radial inflow speed, 
we should avoid configurations that produce the
tangent point on or near the disk midplane.
Hence,
inclined disks should be used, 
and $\alpha\approx0$\deg\ sightlines should be avoided.

Since we have observationally deduced 
how the disk tilts for our galaxies,
we use the asymmetry between the tilt and the LOS velocity range
to constrain the inflow speed or reject the detection of radial inflow.
But even if the disk tilt is not measured \textit{a priori},
we can use the asymmetry to make 
a testable prediction on which way the disk tilts
if only one of them can reproduce the measured LOS velocity range.  
For example, 
the radial inflow modeling for J142459+382113 
would predict a positive disk inclination. 
But because we deduce a negative inclination 
from the rotation curve and the wrapping direction of the spiral arms,
we can then conclude that the sightline does not detect 
radial inflow on the disk plane.

\subsection{Radial Inflow Detection and Inside-out Disk Growth}
\label{ssec:inside_out}

A galaxy disk with a bluer outer disk but a redder inner disk 
indicates that the outer disk has 
a younger stellar population 
\citep{deJong1996,Bell2000,MacArthur2004}.  
While dust can contribute to the color gradient,
it is unlikely that 
the gradient, especially at the outer disk, 
is largely due to dust extinction
(Section 5 of \citealt{MacArthur2004}).
Hence, a bluer outer disk suggests that 
the disk grows inside-out.
For J142459+382113 and J102907+421737, 
we have obtained images in all F390W, F814W, and $K_s$ bands,
and the color images (Figure~\ref{fig:colorim})
show the presence of the color gradients.  
Hence, both disks need a gas supply 
to grow the disks and 
fuel the star formation in the outer disks.

However, for J142459+382113,
neither a rotating disk  
nor the radial inflow model 
can explain the broad velocity range 
spanned by the \mgII\ absorption.
This suggests that the sightline
does not detect radial inflow near the disk plane.  
A disk growing inside-out but without radial inflow detected
may seem contradictory. 
However,
inflow may not need to be axisymmetric,
such as the infalling streams seen in simulations 
\citep[e.g.,][]{Nelson2013,Stewart2017}.  
Our sightline may simply miss the inflow: 
both observations and simulations suggest that
inflows have a small covering factor.
Down-the-barrel spectral  observations 
only identify inflows in a few percent of galaxies
\citep{Martin2012,Rubin2012},
a result consistent with inflow covering a small solid angle, 
and thereby agrees with simulations that predict
a small covering factor for cold accretion streams
(e.g., \citealt{FaucherGiguere2011,vandeVoort2012}).
Therefore,
whether or not 
we detect radial inflow near the disk plane 
does not imply the presence or absence of
a disk color gradient or inside-out disk growth.  
Instead, our result possibly informs us 
about the spatial geometry of the inflowing gas.

On the other hand, for J102907+421737, 
we have constrained the maximum radial inflow speeds
as 40\kms; hence,
the inflow possibly feeds the growing disk.
We are also able to constrain the radial inflow speeds
for another galaxy, J091954+291345, as 30\kms. 
A 30-40\kms\ radial inflow speed is consistent with 
inflow gas measurements and models 
of the Milky Way and nearby galaxies.  
For example,
Complex C falls towards the Galactic plane
at 50-100\kms\ \citep{Wakker1999}, and
the extraplanar \hI\ gas in the Milky Way 
has shown evidence of infall motion at 20-30\kms\citep{Marasco2011}.
As for nearby galaxies, 
the \hI\ gas of NGC 2403 moves radially inward at 
10-20\kms\ \citep{Fraternali2002}, 
and the \siIV\ absorbing gas in M33 has 
a modeled vertical accretion speed of 110\kms\ 
at the disk-halo interface
\citep{Zheng2017}.
In our model, 
we have only considered infall in the radial direction
on the disk plane 
but not infall perpendicular to the disk plane.
Since our sightlines have impact parameters of 
tens of kpc instead of a few kpc as in \citet{Zheng2017},
it becomes unphysical if 
the gas falls towards the extended disk plane at large radii 
but not towards the central region of the galaxy,
where the gravitational potential is the lowest. 
And because of  the corotation between 
the \mgII\ gas and the disk rotation 
(\citetalias{Ho2017} and \citetalias{Martin2019})
and the formation of extended gas disks in simulations
(see Introduction),
we have modeled radial inflow in the disk plane
instead of spherical inflow.  
We emphasize, however, 
that although the gas kinematics require the CGM to have
a component of angular momentum aligned with 
that of the disk,
they do not directly constrain the location of 
the absorbing gas along the LOS.  
Radial inflow in the disk plane is 
consistent with these constraints, 
but other solutions may be possible, 
e.g., satellites accreting onto the galaxy
\citep{Shao2018}, 
series of gas clouds intercepted by the sightline
but not all clouds are spiraling toward the galaxy, etc.
Our simple radial inflow model predicts inflow speeds 
that are at least consistent with measurements 
for the Milky Way and nearby galaxies.

\section{Conclusion}
\label{sec:conclusion}

In this paper, we determined 
the 3D orientation of galactic disks 
to constrain circumgalactic gas flow models.  
We presented new images from 
Keck/NIRC2 (with LGSAO) and HST/WFC3 to
determine the orientation of galaxy disks on the sky.
Since spiral arms trail the direction of rotation,
the rotation curve measurements 
and the wrapping direction of spiral arms 
reveal the disk orientation in 3D-space, 
i.e., the 3D disk orientation
(Figure~\ref{fig:rc_arm_flip} and \ref{fig:galim}).

We modeled the \mgII\ gas kinematics 
measured at four quasar sightlines 
around $z\approx0.2$, star-forming galaxies.
We selected the galaxy--quasar pairs 
from \citetalias{Ho2017} and \citetalias{Martin2019},
and all sightlines intersect 
the inner CGM and lie within 30\deg\ from 
the galaxy major axes. 
We used the new galaxy images and the 
rotation curve measurements to determine
the 3D disk orientation of these galaxies.
Together with our inflow model
with gas spiraling inward in the disk plane, 
we modeled the broad LOS velocity range 
spanned by the \mgII\ absorption,
for which the broad velocity ranges
are known to pose challenges in kinematic modeling.

Combining our radial inflow model and the measured disk tilt
can constrain the inflow speed or 
rule out radial inflow detection.
We constrained the 
maximum radial inflow speeds 
for two galaxies as 30-40\kms,
consistent with 
inflow measurements for the Milky Way 
and nearby galaxies.  
On the other hand, we excluded the detection of 
radial inflowing gas in the sightline of one galaxy,
because the disk tilts in a direction 
opposite to that permitted by the model.

The inflow speeds constrained by our model
are also observed in simulations.
The inferred inflow speeds agree with those
for galaxies of similar masses
in the \texttt{EAGLE} simulations;
the comparable mass inflow rates (order of unity in \msunyr) 
and galaxy star formation rates 
suggest that the inflowing gas plausibly
sustains the galaxy star-forming activities \citep{Ho2019},
and the gas precesses 
and aligns its angular momentum vector with 
the pre-existing cold gas in the galactic disk
\citep{Stevens2017}.
Low-redshift, Milky Way-like galaxies in 
the \texttt{FIRE} simulations also 
show order-of-unity mass inflow rates in total,
which include the contribution from 
distinct sources of materials growing the galaxy,
e.g., 
fresh accretion from the intergalactic medium, 
accretion of wind materials expelled by the galaxies themselves
and/or the gas exchange between galaxies via winds
\citep{Muratov2015,AnglesAlcazar2017}.

The bluer outer disks of our galaxies indicate
that the outer disks are younger, 
consistent with disks growing inside-out.  
This may seem at odd with 
not detecting radial inflow for the one galaxy.  
However, 
the galaxy disk may accrete gas from infalling streams,
but our sightline has not fortuitously intersect the streams
due to their small covering factor.
The ``non-detection'' thereby informs us 
about the spatial geometry and the distribution 
of the feeding gas.

One of our examples demonstrated
that the disk inclination angle
and the azimuthal angle of the quasar sightline
also affect whether 
the disk tilt can effectively constrain the radial inflow speed.  
Typically, we can constrain or rule out radial inflow
because flipping the disk changes how the 
radial and rotation velocity vectors add together,
producing different LOS velocities and velocity ranges.
However, for a less inclined disk, 
the projected LOS velocity becomes insensitive to the 
variation of the radial inflow speed.
And for a sightline at small azimuthal angle,
because the tangent point
lies near the disk midplane, 
the two disk tilts produce 
LOS velocities that are hardly distinguishable.
Hence, 
in the future, 
to search for and model 
radial inflow on the galactic disk plane,
observers should 
select inclined disks and 
avoid $\alpha\approx0$\deg\ sightlines.


\acknowledgments

We thank the referee for 
the thoughtful comments and suggestions 
that improved the organization of the manuscript.  
This research was supported by 
Space Telescope Science Institute 
under grant HST-GO-14754.001-A
and the National Science Foundation under AST-1817125.
The rotation curve data used here 
were obtained 
at the Apache Point Observatory (APO) 3.5-meter telescope, 
which is owned and operated by the Astrophysical Research Consortium.
Observing time was allocated to 
the New Mexico State University through 
Chris Churchill, who is supported by 
the National Science Foundation under AST-1517816.
These APO data were previously published
in \citetalias{Ho2017} and \citetalias{Martin2019},
and we gratefully 
acknowledge the contributions of Chris Churchill and Glenn Kacprzak
to the rotation curve data.
Some of the data used herein were obtained 
at the W. M. Keck Observatory, 
which is operated as a scientific partnership 
among the California Institute of Technology, 
the University of California and 
the National Aeronautics and Space Administration. 
The Observatory was made possible by the 
generous financial support of the W. M. Keck Foundation. 
The authors wish to recognize and acknowledge 
the very significant cultural role and reverence 
that the summit of Maunakea has always had within 
the indigenous Hawaiian community.  
We are most fortunate to have the opportunity 
to conduct observations from this mountain.

\facilities{HST (WFC3), Keck:II (NIRC2), Keck:I (LRIS), ARC (DIS)}

\bibliography{master_im}

\end{document}